\documentclass[pdflatex,sn-vancouver,Numbered]{sn-jnl}
\usepackage[numbers,sort&compress]{natbib}
\bibpunct{(}{)}{,}{n}{}{}

\usepackage{graphicx}%
\usepackage{multirow}%
\usepackage{amsmath,amssymb,amsfonts}%
\usepackage{amsthm}%
\usepackage{mathrsfs}%
\usepackage[title]{appendix}%
\usepackage{xcolor}%
\usepackage{textcomp}%
\usepackage{manyfoot}%
\usepackage{booktabs}%
\usepackage{algorithm}%
\usepackage{algorithmicx}%
\usepackage{algpseudocode}%
\usepackage{listings}%
\usepackage{float}
\usepackage{bm}
\usepackage{adjustbox}
\usepackage{newtxtext,newtxmath}
\usepackage{titlesec}
\usepackage{xcolor}

\definecolor{DarkBlueText2}{RGB}{14,40,100}

\makeatletter
\renewcommand\section{\@startsection{section}{1}{\z@}%
   {-3.5ex \@plus -1ex \@minus -.2ex}%
   {2.3ex \@plus.2ex}%
   {\normalfont\Large\bfseries\color{DarkBlueText2}}}

\renewcommand\subsection{\@startsection{subsection}{2}{\z@}%
   {-3.25ex\@plus -1ex \@minus -.2ex}%
   {1.5ex \@plus .2ex}%
   {\normalfont\large\bfseries\color{DarkBlueText2}}}

\renewcommand\subsubsection{\@startsection{subsubsection}{3}{\z@}%
   {-3.25ex\@plus -1ex \@minus -.2ex}%
   {1.5ex \@plus .2ex}%
   {\normalfont\normalsize\bfseries\color{DarkBlueText2}}}
\makeatother

\usepackage{lineno}
\theoremstyle{thmstyleone}%
\theoremstyle{thmstyletwo}%

\theoremstyle{thmstylethree}%

\begin{document}

\makeatletter
\renewcommand{\ps@headings}{%
  \renewcommand{\@oddhead}{\hfil\thepage\hfil}%
  \renewcommand{\@evenhead}{\hfil\thepage\hfil}%
  \renewcommand{\@oddfoot}{}%
  \renewcommand{\@evenfoot}{}%
}
\pagestyle{headings}
\makeatother

\title[]{Intramuscular microelectrode arrays enable highly-accurate neural decoding of hand movements}

\author[1]{\fnm{Agnese} \sur{Grison}}%\email{agnese.grison16@imperial.ac.uk}
\author[2,3]{\fnm{Jaime} \sur{Ib\'{a}\~{n}ez Pereda}}%\email{jibanez@unizar.es}
\author[1,4]{\fnm{Silvia} \sur{Muceli}} %\email{muceli@chalmers.se}
\author[1]{\fnm{Aritra} \sur{Kundu}}%\email{a.kundu@imperial.ac.uk}
\author[5]{\fnm{Farah} \sur{Baracat}}%\email{farah.baracat@uzh.ch}
\author[5]{\fnm{Giacomo}\sur{Indiveri}}
\author[5]{\fnm{Elisa}\sur{Donati}}
\author*[1]{\fnm{Dario} \sur{Farina}}\email{d.farina@imperial.ac.uk}

\affil[1]{\orgdiv{Department of Bioengineering}, \orgname{Imperial College London}, \orgaddress{\city{London}, \country{United Kingdom}}}

\affil[2]{\orgdiv{BSICoS group, I3A Institute and IIS Aragón}, \orgname{University of Zaragoza}, \orgaddress{\city{Zaragoza}, \country{Spain}}}

\affil[3]{\orgdiv{Centro de Investigacion Biomedica en Red en Bioingeniera, Biomateriales y Nanomedicina}, \orgname{CIBER}, \orgaddress{\city{Zaragoza}, \country{Spain}}}

\affil[4]{\orgdiv{Department of Electrical Engineering}, \orgname{Chalmers University of Technology}, \orgaddress{ \city{Gothenburg}, \country{Sweden}}}
\affil[5]{\orgdiv{Institute of Neuroinformatics}, \orgname{University of Zurich and ETH Zurich}, \orgaddress{ \city{Zurich}, \country{Switzerland}}}

\abstract{Decoding the activity of the nervous system is a critical challenge in neuroscience and neural interfacing. In this study, we present a neuromuscular recording system that enables large-scale sampling of muscle activity using microelectrode arrays with over 100 channels embedded in forearm muscles. These arrays captured intramuscular high-density signals that were decoded into patterns of activation of spinal motoneurons. In two healthy participants, we recorded high-density intramuscular activity during single- and multi-digit contractions, revealing distinct motoneuron recruitment patterns specific to each task. Based on these patterns, we achieved perfect classification accuracy (100\%) for 12 single- and multi-digit tasks and over 96\% accuracy for up to 16 tasks, significantly outperforming state-of-the-art EMG classification methods. This intramuscular high-density system and classification method represent an advancement in neural interfacing, with the potential to improve human-computer interaction and the control of assistive technologies, particularly for replacing or restoring impaired motor function.}

\keywords{motoneuron, motor control, intramuscular EMG, human-machine interfaces}

\maketitle

\section{Introduction}

% Situation
Neural interface research aims to restore or establish connections between the nervous system and the external environment, for example, in individuals with neurological impairments~\cite{hatsopoulos2009science, lebedev2006brain}. These interfaces, based on stimulating (encoding) and/or recording (decoding) of neural pathways, are used to assist individuals with disabilities or enhance the capabilities of healthy individuals. For this purpose, the decoding of neural signals provides commands to virtual systems, wearable robots, or tele-operated devices. 

Direct brain signal decoding is the only option for severe impairments; however, for many other applications, nerve or muscle interfacing may be preferable~\cite{farina2014bionic, oskoei2007myoelectric}. The surface electromyogram (EMG) is particularly simple to record and is used in several interfacing applications~\cite{zheng2022surface, shenoy2008online, wolczowski2010human}. Although recording signals from muscle tissue is not a direct neural interface, the EMG captures the electrical activity produced by muscles, which reflects the neural activation sent from the spinal cord's output layers to the muscles~\cite{farina2017man}. Therefore, surface EMG signals can be decoded to reflect the activities of individual spinal motoneurons, which can then be mapped into estimated motor intent.

Despite the potential of surface EMG-based decoding of neural activity, current decoding and mapping systems are limited in accuracy and, particularly, vary significantly in performance across individuals and conditions. The primary cause of this variability is the dependence of surface EMG characteristics on the volume conductor separating the sources of signal (muscle fibers) and the recording electrodes. Associated with this limitation are the sensitivity of surface EMG features to electrode misplacement, typically due to donning and doffing of the wearable device~\cite{mesin2009surface}, and EMG signal cross-talk~\cite{muceli2013extracting, farina2002surface}. These and other limitations of wearable EMG sensors can be overcome with invasive electrodes. Recent advances in intramuscular EMG technology have provided microelectrode arrays with tens of electrodes closely spaced with respect to each other (high-density EMG microelectrode arrays)~\cite{muceli2015, muceli2022blind}.

\begin{figure}
    \centering
    \includegraphics[width=\linewidth]{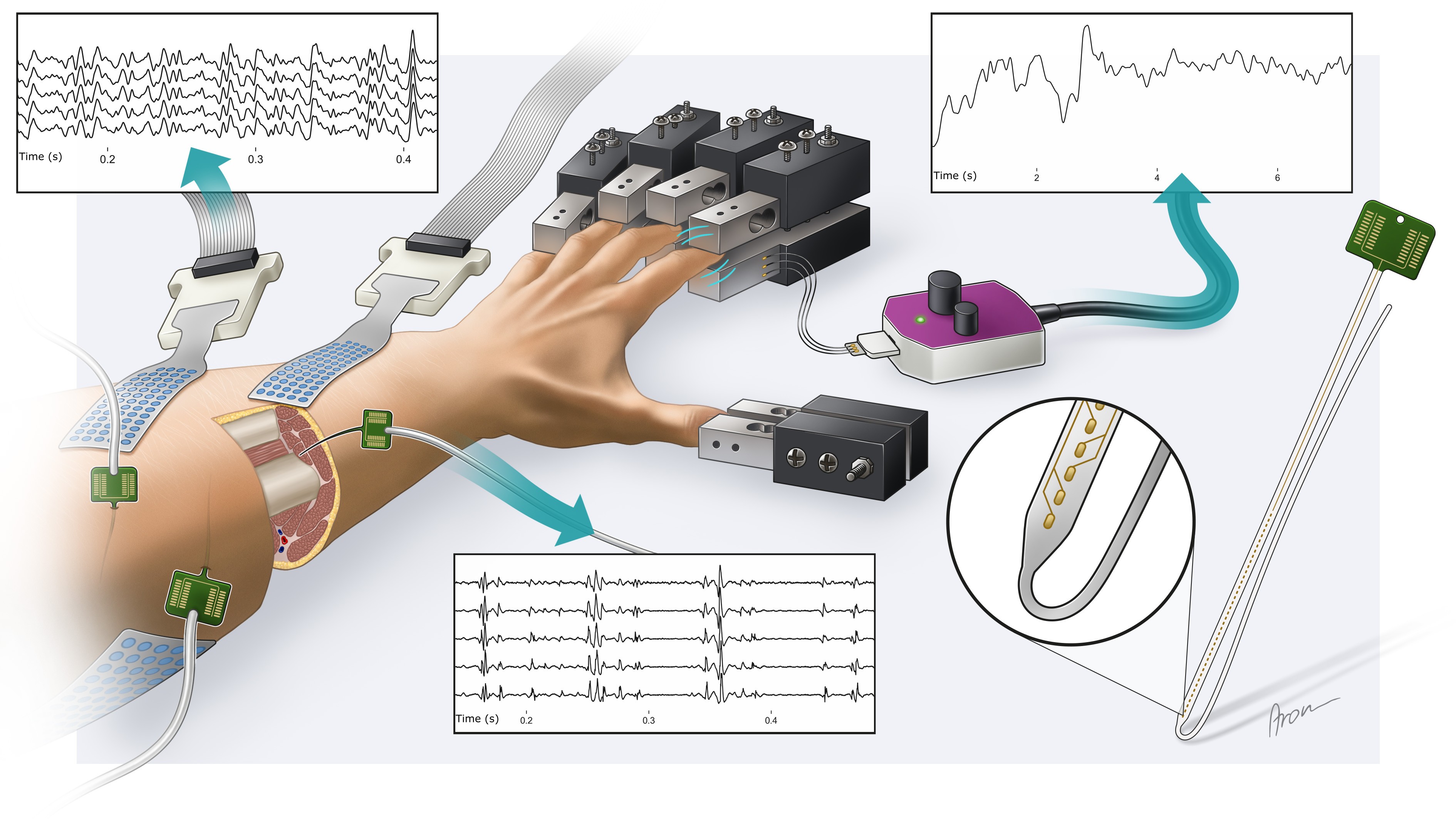}
    \caption{\textbf{Experimental setup and intramuscular high-density EMG recordings.}  Three microelectrode arrays were implanted into the forearm muscles of two participants and three high-density surface electrode grids were placed on the skin overlying the intramuscular recording sites. Two load cells per finger were used to measure flexion and extension forces. Representative data are shown for intramuscular EMG, surface EMG, and force recordings. A schematic of the high-density intramuscular microelectrode arrays, with a detailed zoom-in on the electrode tip, is shown on the bottom right.}
    \label{fig: setup}
\end{figure}

Here, we present an innovative neural decoding system that extracts neural information from high-density intramuscular EMG (HD-iEMG) microelectrode array signals, mapping precise discharge timings of the decoded motoneurons into the executed tasks. This system was experimentally validated offline in two healthy individuals, using three concurrently implanted intramuscular electrodes in each participant (Fig. \ref{fig: setup}). The validation involved classifying single- and multi-digit isometric tasks. The performance of HD-iEMG mapping was then compared to the use of traditional pattern recognition techniques applied to high-density surface EMG (HD-sEMG) and HD-iEMG signals. Therefore, we present three key advances in the field of neural interfacing: 1) recordings of 120 intramuscular signals concurrently detected, paired with corresponding HD-sEMG signals. This is so far the largest reported number of implanted electrodes in the forearm muscles; 2) a highly accurate neural classifier (100\% accuracy in single- and multi-digit tasks for 12 classes), using physiological and interpretable features; 3) a comprehensive comparison of mapping performance across invasive and non-invasive high-density recording modalities. These findings enhance our understanding of neuromuscular control of the hand and have significant implications for neuroprosthetics and human-machine interfaces.

\section{Results}\label{sec2}

\subsection{Targeted muscle insertions}
\label{subsubsec:US}
\begin{figure}[!ht]
    \centering
     \includegraphics[width=\textwidth]{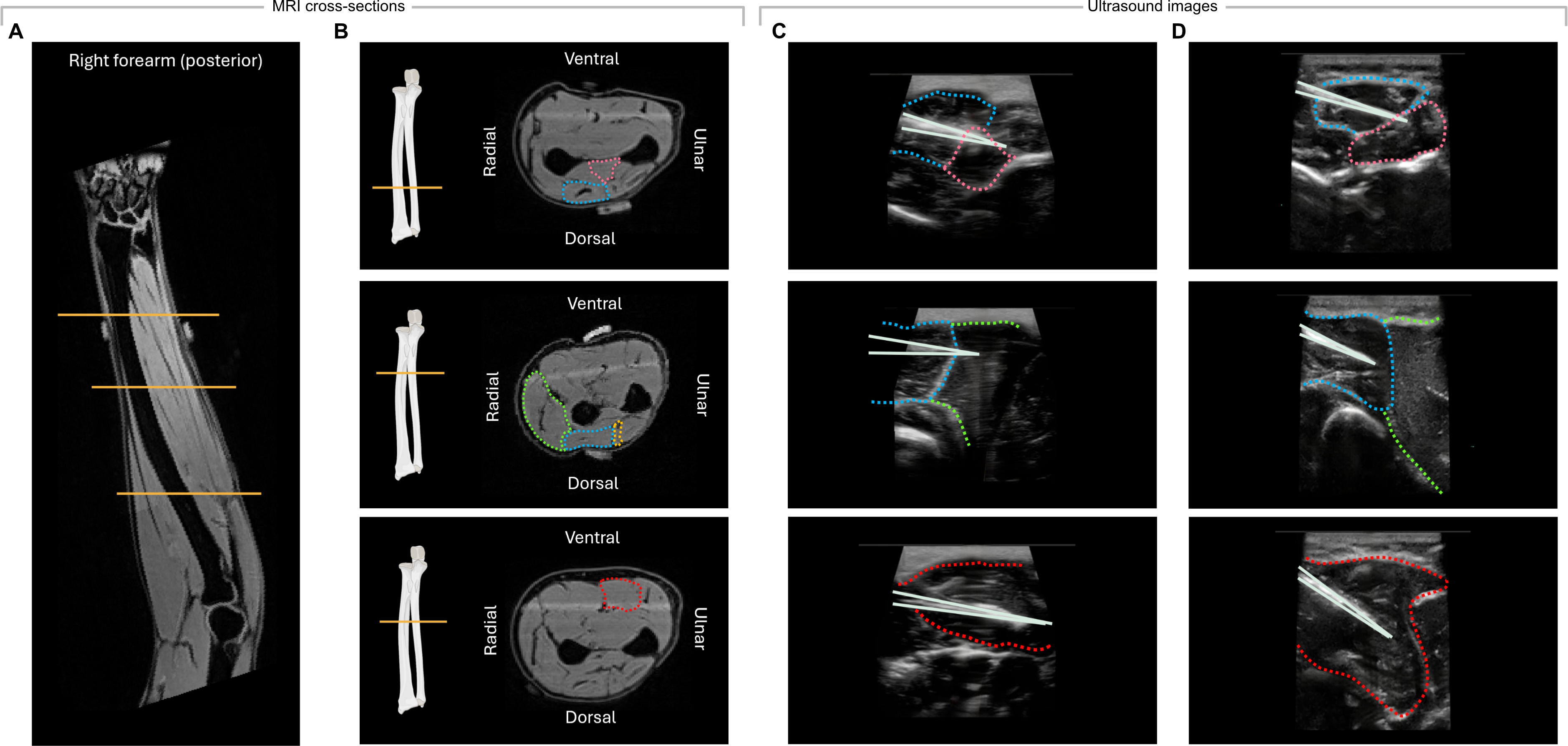}
     \caption{\textbf{Anatomical assessment of forearm muscles and ultrasound-guided insertions}. Muscle boundaries are highlighted in colour (EPL: pink, EDC: blue, EDM: yellow, ECRB: green, FDS: red). \textbf{A} Posterior MRI of the right forearm. Yellow lines mark the levels of each cross-sectional scan in panel B. \textbf{B} Forearm MRI scans from one participant with targeted muscles outlined in colour. The first and second columns include a marker for tracking anatomical landmarks across views. The ulna and radius indicate the forearm level at which each cross-sectional slice was obtained. \textbf{C, D} Ultrasound images during microelectrode array insertion for participant one (\textbf{C}) and participant two (\textbf{D}). The guiding needle is highlighted in white.}
     \label{fig: insertions}
\end{figure}

The recording of multiple high-density intramuscular microelectrode arrays represents one of the key innovations of this study. HD-iEMG enables recordings from a significantly larger number of sites compared to traditional needle or wire electrode arrays~\cite{farina2024neural}. Furthermore, HD-iEMG signals exhibit greater spatial and temporal sparsity than HD-sEMG, allowing for more precise analysis of motoneuron activity~\cite{muceli2022blind}. In this study, three HD-iEMG electrodes were inserted into the forearm muscles of two healthy participants, with one electrode placed in each of the most relevant extrinsic muscles for finger flexion and extension. While several other forearm muscles contribute to these movements, each electrode insertion is technically challenging and time-consuming; therefore, we selected three key muscles to balance feasibility with exhaustive signal acquisition.
Interfacing with motoneurons that control hand function poses inherent challenges due to the precise targeting required for intramuscular electrode placement in the muscles responsible for dexterous control. Optimal electrode placement sites were determined via a magnetic resonance imaging (MRI)- and ultrasound-based anatomical assessment of the forearm. MRI was employed to gain a detailed understanding of the forearm muscle anatomy and to guide the insertion point and angle of the needles. Specific forearm points, based on anatomical landmarks, were tracked using MRI markers across scans, allowing precise localisation of marker positions and detailed insights into forearm structure (Fig. \ref{fig: insertions} A, B). Identification of the muscles with the ultrasound was aided by having participants performing voluntary contractions of the targeted muscles. Each electrode array, comprising 40 platinum electrodes (140 $\mu m$ $\times$ 40 $\mu m$) linearly distributed over 2 cm with 0.5 mm inter-electrode (IED) spacing~\cite{muceli2022blind}, targeted muscles involved in finger flexion and extension. The insertion of each array was facilitated by a 25-gauge hypodermic needle, which was attached to the electrode with a guiding filament. Once the electrode was positioned, the filament was cut, allowing the needle to be removed while the electrode remained securely in the muscle~\cite{muceli2022blind}.

The first microelectrode array targeted the extensor pollicis longus (EPL) and extensor digitorum communis (EDC) muscles, activated for the extension of the thumb and the digits, respectively. A distal location in the forearm was chosen to allow concurrent monitoring of the two muscles with the same microelectrode. The second array was inserted transversely to target the extensor digiti minimi (EDM) and the extensor digitorum communis (EDC), employing an unconventional technique that maximises coverage of the muscles relevant to digit extension. The third array targeted the flexor digitorum superficialis (FDS) muscle, chosen for its accessibility and essential role in flexion tasks. 

MRI provides a comprehensive anatomical view of the forearm, but practical constraints limit its use during needle insertions. Once the electrode insertion points and general positioning were identified using MRI, precise muscle targeting during the actual insertion was critical to avoid veins, arteries, and nerves. This necessitated real-time imaging guidance to ensure accurate and safe insertion. A portable ultrasound device (Butterfly iQ+) was used to visualise the electrode paths to the target muscles in real-time (Fig. \ref{fig: insertions} C and D, for Participant 1 and Participant 2, respectively). Since the EDC and FDS have multiple compartments supplying different fingers, it is difficult to determine the exact location being sampled; however, we attempted to target all compartments by instructing participants to perform voluntary movements and verifying muscle activation. Post-insertion, participants performed maximal muscle contractions (MVC) as a reference for submaximal tasks. This procedure also confirmed secure electrode placement and minimised the risk of displacement.

While electrode placement followed the above rigorous procedure, some errors in the exact target placements occurred. For example, the limited field of view of the ultrasound probe impeded the full tracking of the microelectrode array targeting the EDM for both participants (Fig. \ref{fig: insertions} C, D, second row). The extension of the insertion to the boundary of the Extensor Carpi Radialis Brevis (ECRB) suggests that the electrode might not have fully targeted the EDM, potentially failing to record specific information related to the little finger tasks. Moreover, although the EPL was targeted in both participants (Fig. \ref{fig: insertions} C, D, first row), the number of channels within the muscle was relatively small. This potentially impacted the amount of information recorded from the EPL during these tasks.

\subsection{Data characteristics}
Concurrent recordings of HD-iEMG and HD-sEMG were obtained during flexion and extension tasks for each finger of the hand individually and for four combinations of finger tasks. Two repetitions were recorded for each task. By decoding the electrical activity into the spiking activities of spinal motoneurons, we performed classification analyses across multiple conditions. First, we classified 16 classes of tasks, which included all individual finger contractions and three two-finger combinations, in both flexion and extension. Next, we analysed the 12 classes that corresponded to tasks primarily controlled by the implanted muscles, ensuring that these tasks involved contractions whose neural activity was captured by the intramuscular microelectrode arrays. Finally, as a proof of concept for prosthetic hand control (one representative application of this mapping system), we selected eight functionally relevant tasks, which are challenging to decode with current methods (see Section \ref{methods: experiments} in Methods for more details).

\begin{figure}[!ht]
    \includegraphics[width=\textwidth]{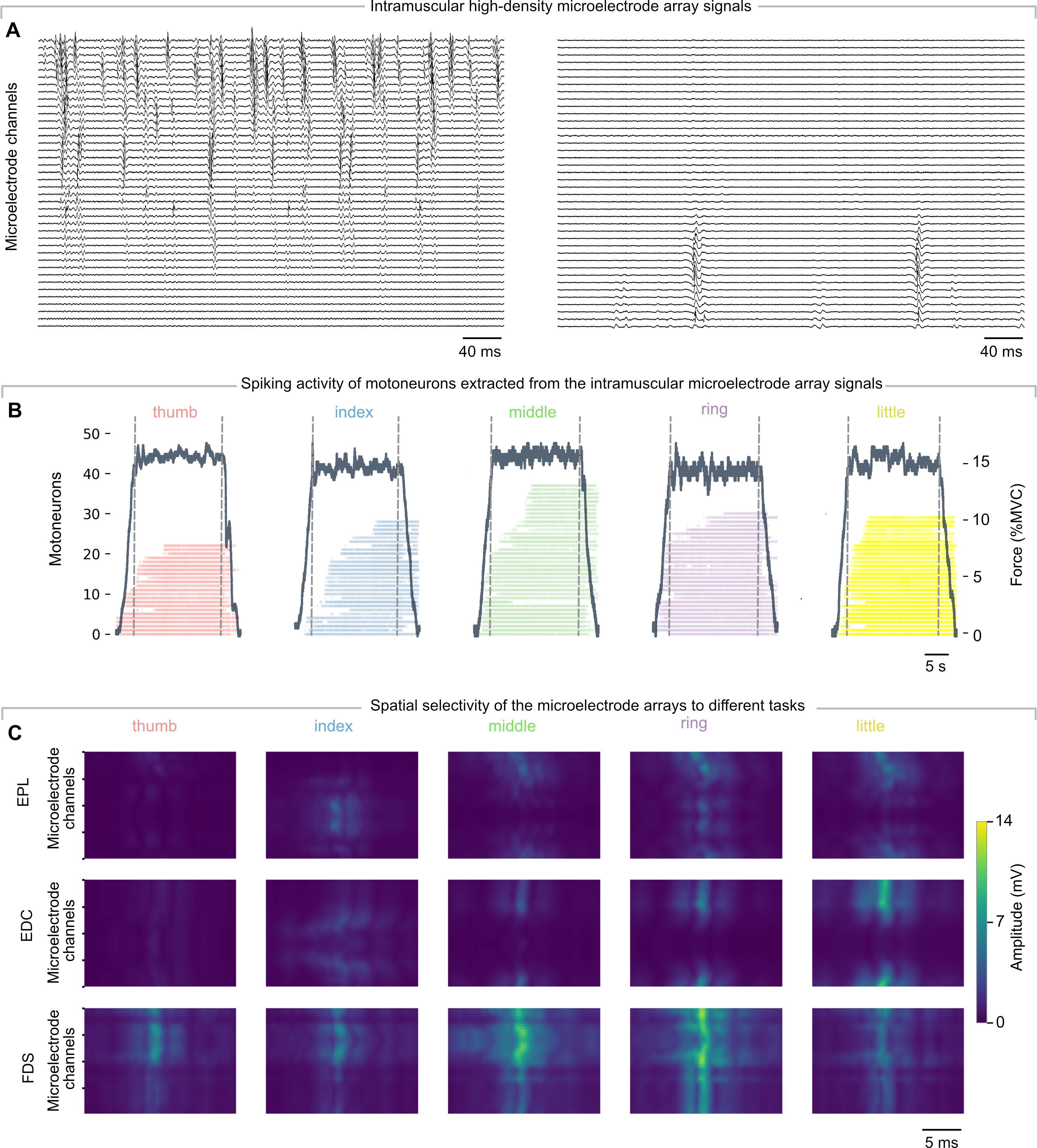}
        \caption{\textbf{Data characteristics}. \textbf{A} A 0.5-second segment of HD-iEMG data captured during the extension from the EPL electrode of the little finger (left) and of the ring finger (right). \textbf{B} Raster plots showing the neural activity extracted from the extension contractions of individual fingers for the second repetition for participant one. The sections within the dashed lines indicate the parts of the signal considered for this study. The y-axis for the motoneurons is not shared across plots. \textbf{C} Summed amplitude map of motoneuron activities during the extension task of each finger and for each electrode (participant one).}
        \label{fig: data}
\end{figure}

Figure \ref{fig: data}A presents a representative example of signals recorded from one of the intramuscular microelectrode arrays during two different extension tasks, illustrating their spatial distribution across the array. The multi-channel data recorded from all three microelectrode arrays was independently decomposed to identify the spiking times of individual motoneurons (Fig. \ref{fig: classification_pipeline}A, \ref{fig: data}B). On average, 27.1±12.9 (participant one) and 8.0±6.1 (participant two) motoneurons were identified from the HD-iEMG per task (mean and std across two repetitions, the three electrodes, and the 16 tasks). The average discharge rate of the decomposed motoneurons across the two participants was 11.0±3.9 pulses/s. In contrast, the HD-sEMG recordings yielded 5.1±4.4 (participant one) and 1.8±1.8 (participant two) motoneurons on average. The average discharge rate of the decomposed motoneurons across the two participants was 8.8±3.1 pulses/s. Due to the low number of motoneurons identified from the surface recordings, the subsequent analysis was limited to motoneurons identified from the intramuscular recordings.
Figure \ref{fig: data}C presents the summed amplitude maps of motoneuron activation during different finger extension tasks. These maps were generated by first computing the spike-triggered average for each motoneuron action potential, per electrode, and per task, followed by normalisation and rectification. The spike-triggered averages were then summed across all active motoneurons for each electrode and task to highlight the areas of the electrode array that were most active for each task. As shown in Figure \ref{fig: data}C, the electrodes exhibited task-specific activity, indicating localised activation within each array. This observation reflects the high spatial selectivity of each HD-iEMG channel. 
Notably, the EPL exhibited only slight activity during thumb extension but was more active during the extension of the other digits. This pattern aligns with expectations (see Fig. \ref{fig: insertions}), as the EPL electrode placement did not optimally capture the targeted muscle. In contrast, the EDC was selectively active during the extension of all digits except the thumb, consistent with its anatomical function as the primary extensor of the index, middle, ring, and little fingers. Additionally, substantial co-contraction of the flexor muscles was observed during digit extension, a phenomenon that is expected to stabilise the wrist and fingers during the contraction. This was particularly evident in the FDS, which exhibited higher activation, likely due to its large muscle volume and functional role in flexion and stabilisation. While some improvements in insertion procedures may be needed in future studies, the results indicated that the recorded information was of high quality for the subsequent analyses.

\subsection{Decoding hand function}

\begin{figure}[!htbp]
    \centering
    \includegraphics[width=\textwidth]{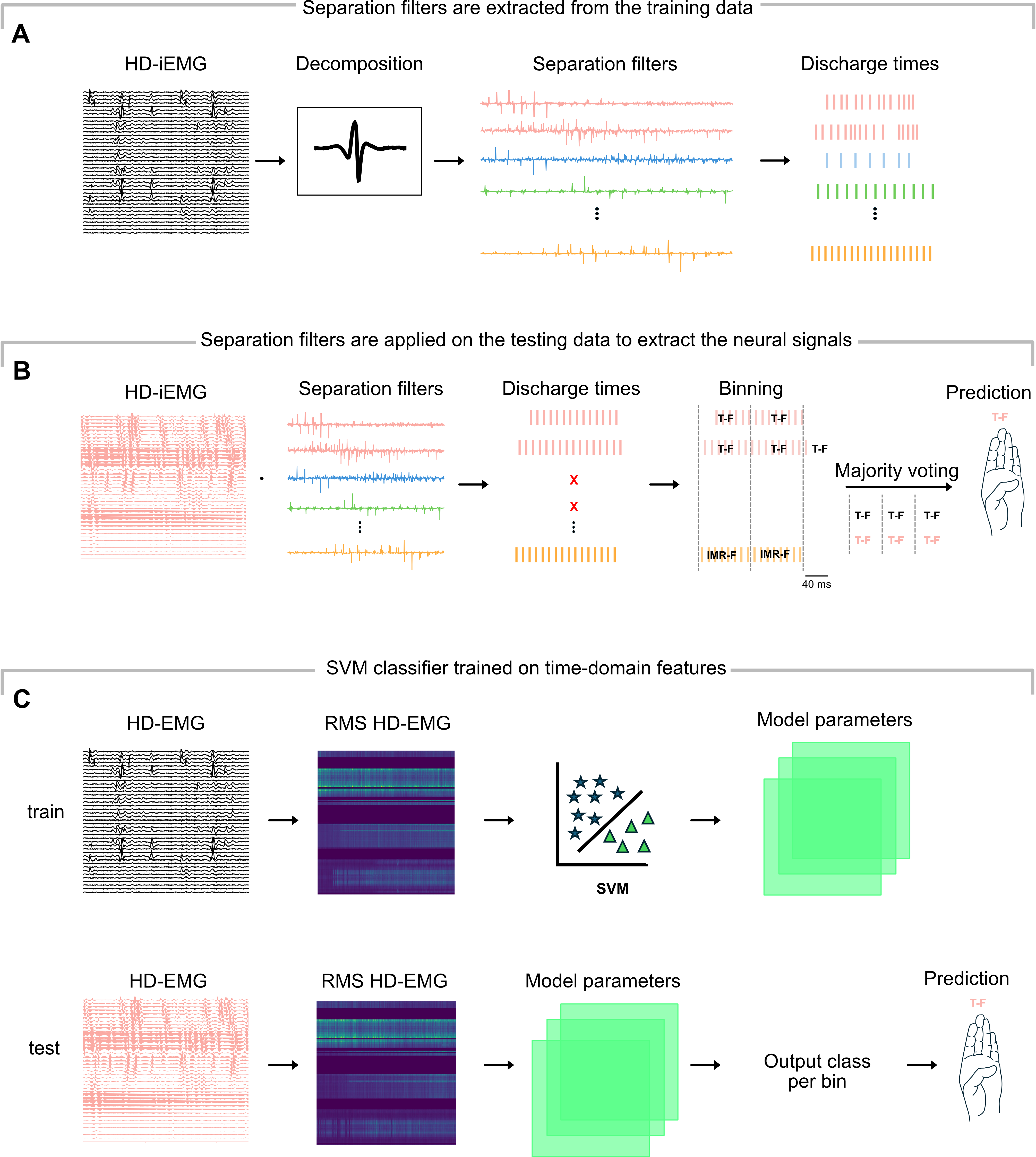}
    \caption{\textbf{Classification pipelines.} \textbf{A} The HD-iEMG signals recorded by each electrode and task in the first recorded repetition were independently decomposed into their constituent neural activities. The separation filters optimised in the decomposition to extract the discharge times were pooled into a dictionary of filters.
    \textbf{B} Classification using the motoneuron discharge times as features. The separation filters were applied to the HD-iEMG signals of each electrode and task, in the second recorded repetition, generating a set of sequences of electrical pulses, also known as Innervation Pulse Trains (IPTs). Each IPT was clustered into source and noise components. If the clustering quality was robust, the source was accepted and segmented into sliding windows. Each time bin was assigned a class label based on the separation filter that produced the corresponding source. Majority voting was applied across the set of IPTs for each segment. The final classification label was determined by the ratio of correctly labelled segments to the total number of labelled segments. \textbf{C} Classification from global EMG features. The RMS values of the HD-EMG signals were extracted and segmented into sliding windows. These formed the input to a linear C-Support Vector Machine (SVM) classifier. The model was fit on the training data (repetition one) and then applied to the test data (repetition two) for classification. The resulting output provided the final classification labels.}
    \label{fig: classification_pipeline}
\end{figure}

\begin{figure}[!htbp]
    \centering
    \includegraphics[width=\textwidth]{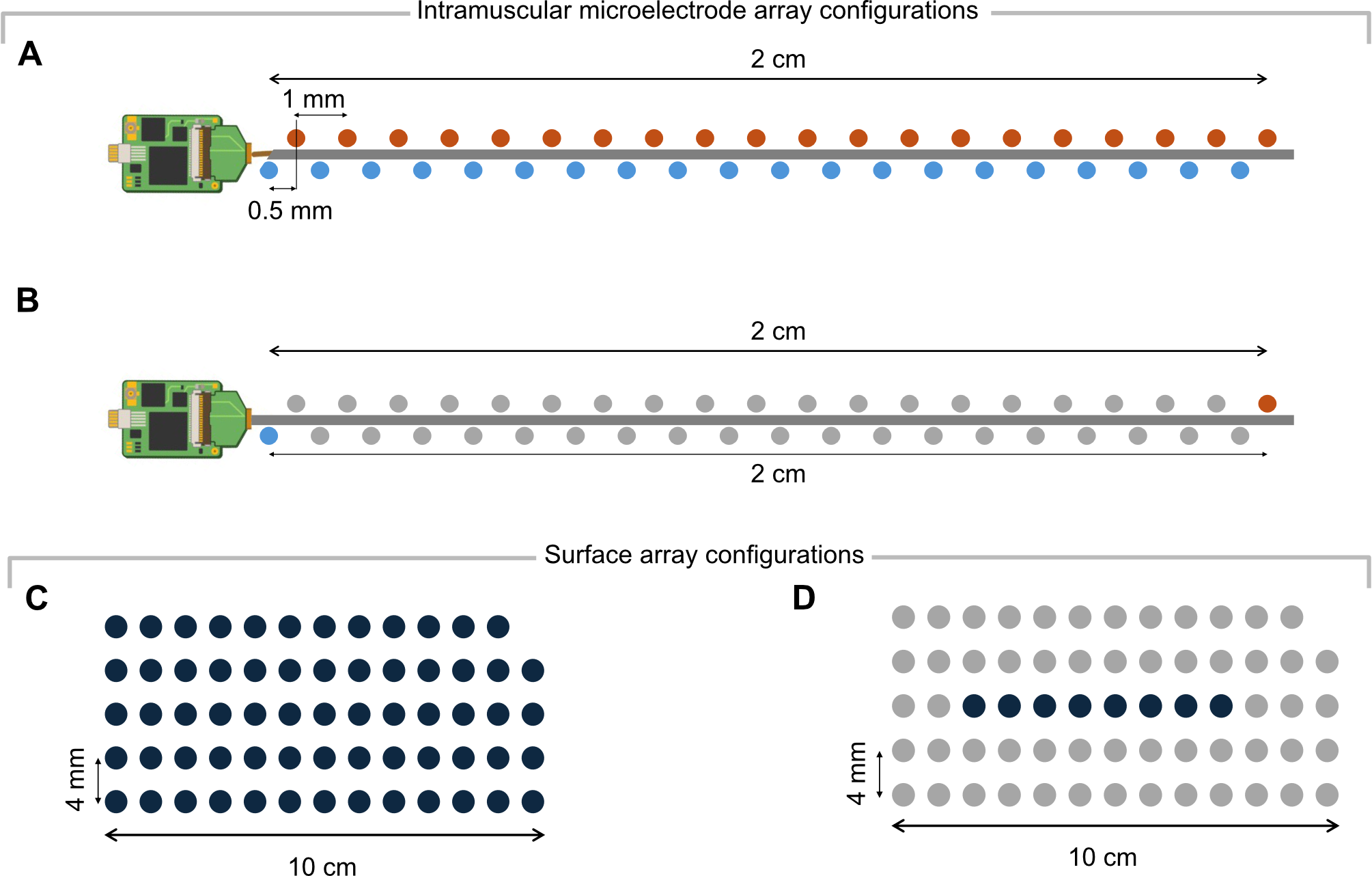}
    \caption{The four electrode configurations considered in this study. \textbf{A} HD-iEMG electrode, 40 channels with 0.5 mm IED. \textbf{B} Lower-density iEMG electrode, two channels with 2 cm IED. \textbf{C} HD-sEMG electrode, 64 channels with 4 mm IED. \textbf{D} Lower-density sEMG electrode, eight channels with 4mm IED.}
    \label{fig: electrode_configs}
\end{figure}

The following sections present the results of three offline analyses: 1) classification of all 16 classes of recorded tasks; 2) classification of a subset of 12 classes associated to tasks primarily requiring the muscles that were targeted by the multi-electrode arrays (see Section \ref{subsubsec:US}), and 3) classification of six (participant one) and eight (participant two) classes chosen as important for prosthetic control~\cite{kashiwakura2023task} (considered as a potential application of the proposed decoding system). In all analyses, results are reported when using three feature spaces: motoneuron discharge timings as decoded from HD-iEMG (Fig. \ref{fig: classification_pipeline}A, B), root mean square (RMS) values for each channel of the HD-iEMG, and RMS values for each channel of the HD-sEMG (Fig. \ref{fig: classification_pipeline}C). Additionally, in the first two analyses, we report results from lower-density iEMG and sEMG signals (configurations visualised in Fig. \ref{fig: electrode_configs}B and D, respectively). This analysis was performed to investigate whether high-density configurations were essential for achieving high prediction accuracy. 

\subsubsection{Full classification (16 classes)} \label{subsec: allfingers}
\begin{figure}[!htbp]
    \includegraphics[width=\textwidth]{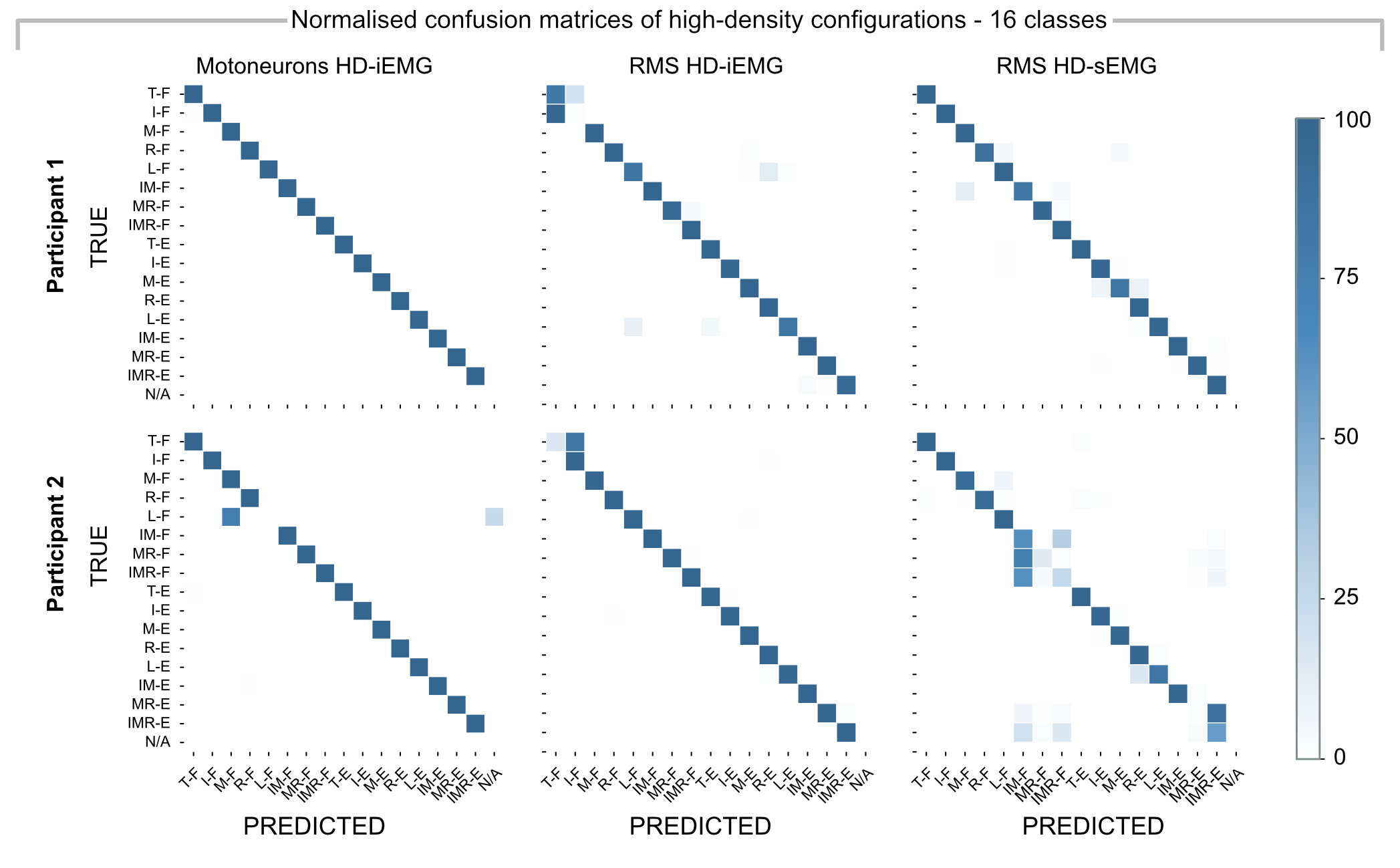}
        \caption{Normalised confusion matrices illustrating the percentage of predictions for each of the 16 flexion and extension tasks, comparing true labels (y-axis) with predicted labels (x-axis). The matrices show predictions for participant one on the first row and participant two on the second row, across different data types: HD-iEMG motoneurons (first column), RMS values of HD-iEMG (second column), RMS values of HD-sEMG (third column).}
        \label{fig: 16classes}
\end{figure}

In the analysis of motor tasks spanning all 16 flexion and extension categories, the motoneuron discharge timings decoded from HD-iEMG provided a perfect classification accuracy for participant one of 100\% and an accuracy of 96.1\% for participant two on average across the tasks. The mis-classification for participant two was exclusively related to the little finger, and it was due to the fact that the electrode placement did not adequately target the EDM, the muscle primarily responsible for the activation of the little finger. 
The accuracy using the decomposition from the HD-iEMG recordings was higher than that achieved using the RMS values of the HD-iEMG (90.1\% and 94.3\%) or the RMS values of the HD-sEMG (96.6\% for participant one, 76.5\% for participant two). Figure \ref{fig: 16classes} presents the confusion matrices per participant and classification type. 

\paragraph{Lower-density electrode configurations}
When using low-density EMG configurations, the accuracy decreased substantially compared to the high-density systems, with the bipolar intramuscular achieving 47.8\% for participant one and 69.0\% for participant two, and the sEMG achieving 92.5\% for participant one and 64.3\% for participant two. These results indicate that the reduction in accuracy when shifting from a high-density to a lower-density configuration for pattern recognition is particularly large for intramuscular recordings and more modest for surface signals~\cite{toledo2019support, bian2017svm}. The confusion matrices for the bipolar and low-density sEMG signals are shown in Appendix \ref{secA1}, Figure \ref{fig: bipolar_surf_16}. 

\subsubsection{Targeted muscle classification (12 classes)} \label{subsec: nothumblittle}

\begin{figure}[!htbp]
    \includegraphics[width=\textwidth]{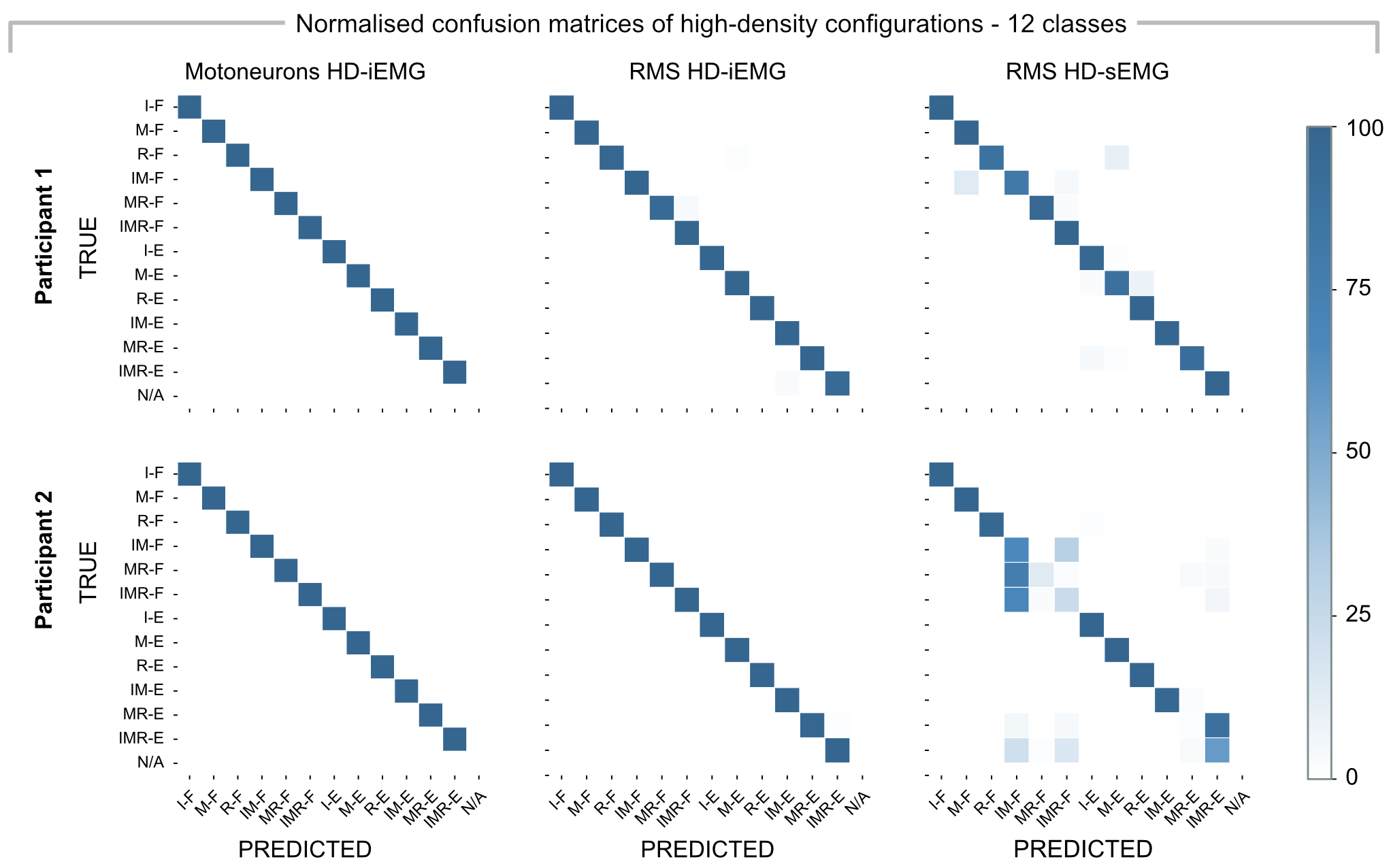}
        \caption{Normalised confusion matrices illustrating the percentage of predictions for each of the 12 flexion and extension tasks, comparing true labels (y-axis) with predicted labels (x-axis). The matrices show predictions for participant one on the first row and participant two on the second row, across different data types: HD-iEMG motoneurons (first column), RMS values of HD-iEMG (second column), RMS values of HD-sEMG (third column).}
        \label{fig: 12classes}
\end{figure}

Figure \ref{fig: 12classes} shows the confusion matrices for the classes fully captured by the electrodes, as confirmed by ultrasound imaging  (Fig. \ref{fig: insertions}). The classifier using motoneurons decomposed from HD-iEMG signals reached 100\% accuracy for both participant one and participant two, which shows a perfect classification when targeting all muscle structures responsible for the tasks. The classifier using RMS values from HD-iEMG achieved an accuracy of 99.3\% for participant one and 99.7\% for participant two, while the classifier using RMS values from HD-sEMG reached an accuracy of 95.3\% for participant one and 71.4\% for participant two. 

These results underscore the superior performance of intramuscular configurations over surface configurations for the classification based on neural features. When the electrodes accurately targeted the relevant muscles, the classification accuracy was 100\% in the two participants.

\paragraph{Lower-density electrode configurations}
On average across the two participants, the bipolar iEMG achieved an accuracy of 55.4\% for participant one and 78.8\% for participant two, while the sEMG achieved an accuracy of 89.1\% for participant one and 57.2\% for participant two. These results show a notable reduction in accuracy compared to high-density configuration systems, highlighting the importance of a high-density setup in achieving high classification accuracy. The confusion matrices for the 12-class classification from the lower-density configurations are shown in the Appendix \ref{secA1}, Figure \ref{fig: bipolar_surf_12}. 

\subsubsection{Mapping specific functional classes (6 and 8 classes)} \label{subsec: interface}
\begin{figure}[!htbp]
    \includegraphics[width=\textwidth]{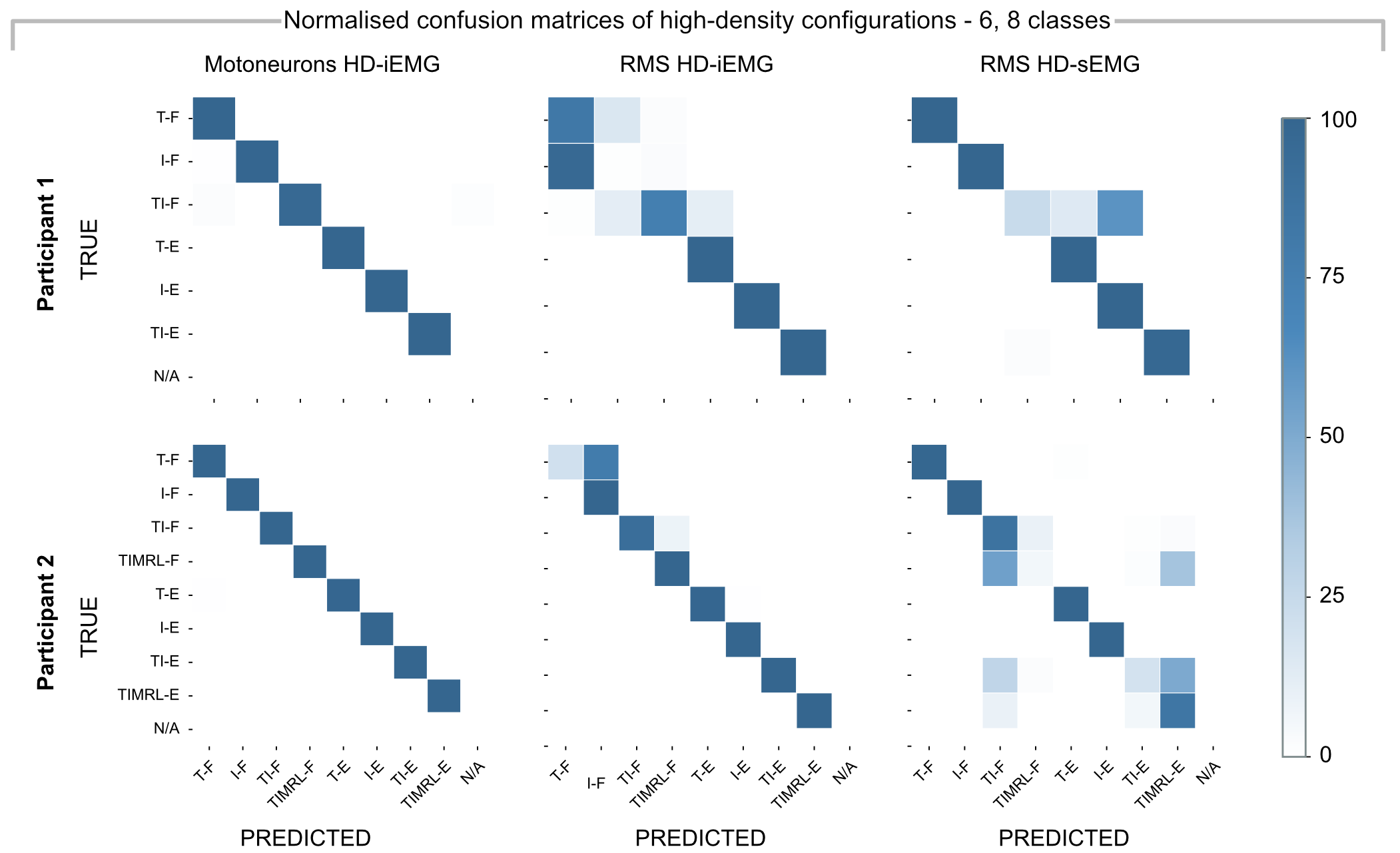}
        \caption{Normalised confusion matrix illustrating the percentage of predictions for each of the six (participant one, first row) and eight (participant two, second row) flexion and extension tasks, comparing true labels (y-axis) with predicted labels (x-axis) across different tasks: motoneurons from the HD-iEMG (first column), RMS values of HD-iEMG (second column), and RMS values of HD-sEMG (third column).}
        \label{fig: interfacing}
\end{figure}

Figure \ref{fig: interfacing} presents the classification results when focusing on the classes most relevant from a control perspective. Results are reported for six (participant one) and eight (participant two) classes (refer to Methods for details). The accuracy achieved from the motoneuron activities decoded from HD-iEMG was 99.4\% for participant one and 100\% for participant two, while the accuracy from the RMS values of the HD-iEMG was 76.4\% for participant one and 89.0\% for participant two. The accuracy from the RMS values of the HD-sEMG was 87.0\% for participant one and 74.4\% for participant two. The effectiveness of using high-density intramuscular EMG recordings combined with motoneuron decomposition for the neural control of prosthesis is also demonstrated in this analysis by the very high accuracy rates (\textgreater99\%). This near-perfect accuracy in classifying tasks typically used for prosthetic hand control indicates the potential of the proposed interface in external device control. 

\section{Discussion}\label{sec12}
Decoding motor intent is critical for accurate control in neural interfacing systems. At the peripheral level, decoding performance depends heavily on the type and quality of recorded muscle signals. We developed an advanced processing system capable of decoding the precise discharge timings of spinal motoneurons by deconvolving HD-iEMG recordings. These recordings capture detailed neural signals from motoneuron pools, offering rich insight into the underlying motor control processes.
The proposed system enabled precise discrimination between hand tasks, which was successfully validated with high accuracy.

Given the high quality of the signals recorded from the intramuscular microelectrode arrays, we hypothesised that a system leveraging the decoded activities of motoneurons could achieve near-perfect accuracy in task classification. 
The accuracy of the proposed mapping system was validated through a series of experiments. A novel signal classification method was applied to the decoded series of discharge timings, achieving near-perfect discrimination for six, eight, and 12 classes (average \textgreater99\%), and very high accuracy for up to 16 classes (average \textgreater96\%). While traditional HD-iEMG and HD-sEMG features demonstrated solid performance (over 71\%), the motoneurons identified through HD-iEMG consistently yielded superior accuracy, exceeding global features by 0.3\% to 28.6\%, depending on the conditions. The substantial improvement in classification accuracy is highly relevant for applications in neuroprosthetics, as this increase in accuracy can significantly enhance the effectiveness and responsiveness of assistive devices, leading to better user experiences and outcomes. Moreover, in all tasks, the classification accuracy from HD-iEMG motoneurons for the combined finger contractions approached 100\% (Figs. \ref{fig: 16classes}, \ref{fig: 12classes}). 

Decoding motor intent can be approached through global EMG features or motoneuron discharge timings, each with distinct advantages and challenges. Global EMG features represent the spatiotemporal convolution of many motoneuron discharges but are variable across individuals, particularly when recorded at the skin surface. This variability, alongside issues like amplitude cancellation and cross-talk, complicates precise control in complex tasks~\cite{farina2004surface, mesin2020crosstalk}. Due to these challenges, global HD-sEMG achieved classification accuracies ranging from 71\% to 97\%, highlighting variability in performance across tasks. Although relatively high classification accuracies were achieved in some tasks, the reliance on features that are not directly linked to the underlying physiological mechanisms diminishes interpretability and presents challenges for clinical applications and motor control research. Understanding the specific contributions of muscle groups and neural activity is essential for validating model predictions and enhancing practical utility. 

In contrast to global EMG features, motoneuron discharge timings provide a direct and physiologically relevant representation of neural activity, closely reflecting the spinal-level coding mechanisms of movement. By decomposing the EMG signals into their constituent motoneuron activities, we demonstrated that motoneuron spike trains can serve as critical sources of information for identifying complex tasks. This approach addresses the variability present in global EMG by separating neural information from muscle fiber action potentials, ensuring that the shapes of action potentials do not influence classification outcomes. Additionally, this separation mitigates the effects of EMG amplitude cancellation and enhances spatial and temporal sparseness, thereby reducing issues related to cross-talk~\cite{farina2014extraction}.

A key novelty of this work is the recording of high-density data intramuscularly, enabling detailed and accurate decoding of neural activity. The importance of intramuscular measurements is underscored by the large number of motoneurons that could be decomposed per task. High-density surface EMG has also been used for motoneuron decoding~\cite{farina2017man} but it presents evident limitations with respect to implanted systems. One of these limitation is that it typically allows for the decoding of fewer motoneurons. This limitation is compounded by the challenge of accurately determining which muscles are activated, primarily due to a lack of spatial selectivity in the signals captured. This lack of resolution can lead to overlapping activity from adjacent muscles, making it difficult to pinpoint specific muscle contributions during movement~\cite{farina2004surface, mesin2020crosstalk}.

Precise targeting of the appropriate muscles and decoding of direct neural information resulted in nearly perfect decoding accuracy from the motoneuron activity derived from HD-iEMG. Ensuring accurate muscle targeting remains critical for effectively extracting useful information from these recordings. As we reported in Section \ref{subsec: nothumblittle}, when there was high confidence in the targeted muscles, the accuracy of classification from decoded motoneuron activities was 100\%. 

Lower-density configurations were extracted from the HD-iEMG and HD-sEMG data to evaluate whether multi-channel configurations were necessary for achieving high classification accuracies. The results showed that classification accuracy from global EMG features dropped with fewer channels, more significantly so for the iEMG. Regarding the 16 class classification, the bipolar iEMG configuration dropped by 40\% for participant one and by 25\% for participant two for 16 classes, while sEMG dropped by 4\% for participant one and just over 10\% for participant two. A similar trend was observed for the 12 classes. This highlights the importance of high-density configurations to accurately classify tasks from the EMG. It has to be noted, however, that in this study, we compared each recording type to only a single low-density configuration, which may not fully generalise to all conditions. On the hand hand, we believe that this provides a representative baseline for evaluating the impact of electrode density on classification performance. Similar results have indeed been obtained by changing the low-density recording in pilot analyses (results not shown).

In a recent study ~\cite{grison2025motor}, we showed that both the number of identifiable motoneurons and the accuracy of their decomposition depend on the number and density of channels in intramuscular microelectrode arrays. Consequently, classification based on motoneuron activity would not be feasible with lower-density configurations, as the decomposition methods used to extract motoneuron signals require high-density setups. For this reason, we did not test decomposition-based features in the case of low-density recordings.

Various approaches have been explored in previous research for classifying hand tasks from global EMG features. For eight classes, time domain and deep learning methods have shown variable level of success, with accuracies ranging from 81.54\%~\cite{fajardo2021emg} to 95\% and about 91\%~\cite{simao2019emg}. A convolutional auto-encoder combined with a CNN achieved 99\% accuracy for 10 finger tasks~\cite{jia2020classification}, but this approach was tested on a smaller set of 10 classes compared to the 16 and 12 classes used in our study.

There have also been previous studies that presented results on the classification of motoneuron activities, extracted from surface EMG, associated to hand tasks. For example, neural drives to muscle regions were estimated by grouping motoneurons based on the spatial location of their corresponding muscle units, achieving over 97\% accuracy across 11 classes in patients with targeted nerve reinnervation~\cite{farina2017man}, while similar methods using motoneuron action potential waveform activations reached around 95\% accuracy for 11 classes~\cite{chen2020hand}. A related technique employing motoneuron filters achieved 94.6\% accuracy for 12 tasks~\cite{chen2023real}. Additionally, neuromorphic approaches have shown promise, with approximately 95\% accuracy for 10 classes~\cite{tanzarella2023neuromorphic}.

Overall, our method outperformed all previous approaches, achieving accuracies exceeding 99\% for six, eight, and 12-class classification tasks and greater than 96\% for 16 classes. Notably, the precision obtained from motoneuron activity significantly surpasses prior results, highlighting the robustness and reliability of our system for decoding neural information. The use of motoneurons not only enhances performance but also provides a theoretical framework for achieving 100\% accuracy. As long as motoneuron discharges are accurately detected, and each task involves at least one unique motoneuron, it becomes theoretically possible to reach perfect classification. This implies that, with precise muscle targeting and continued advancements in decoding algorithms, flawless performance is achievable — a potential we have already demonstrated in some cases in this study (e.g., 100\% accuracy for both participant one and participant two in the 12-class classification). Reaching perfect, or almost perfect, accuracy in decoding is essential, particularly in clinical and real-world applications when even small errors can have significant consequences. For assistive devices, such as prosthetic limb control, near-perfect accuracy is often necessary to ensure reliable, intuitive, and safe operation. Errors in decoding can lead to unintended movements, frustration, or even device rejection by the user. While simpler systems may suffice for some applications, such as general activity recognition, tasks requiring fine motor control demand higher accuracy to maintain usability and effectiveness. The present results bring us substantially closer to this goal. 

% Limitations
Under the experimental conditions of our study, the proposed approach based on decoding motoneurons demonstrated superior accuracy compared to traditional pattern recognition methods. However, the small sample size, driven by the need for intramuscular insertions, limits the generalisation of these results. Further validation with a larger population is needed.

Transitioning to online implementation for real-world applications presents significant challenges. Real-time HD-iEMG decomposition is computationally intensive, requiring high sampling frequencies to capture neural signals accurately~\cite{rossato2023spin}. While advances in microprocessor speeds have made real-time neural data processing more attainable, the challenge remains significant due to the intensive computational demands, especially since even offline data recordings are often hindered by the high sampling rates required (this problem is clearly exacerbated when increasing the number of channels, as in this study). 

A promising future direction is the integration of spiking neural networks directly onto low power mixed-signal neuromorphic chips~\cite{Chicca_etal14, Mead23}. As decomposition methods continue to advance and become more capable of adapting to signal nonstationarities~\cite{guerra2024adaptive}, implementing real-time decomposition methods with neuromorphic electronic circuits will significantly accelerate the development of practical neural interfaces. The combination of adaptive signal processing and the computational efficiency of neuromorphic architectures represents a crucial step toward deploying neural prosthetics and other neurotechnologies that rely on real-time neural decoding~\cite{Donati_Indiveri23}. This integration would not only enhance the system's ability to handle dynamic, real-world scenarios but also reduce the need for constant retraining.

Despite the clear advantages of intramuscular systems over non-invasive ones, practical challenges hinder their widespread adoption. Precise muscle targeting is essential, requiring careful planning and execution by trained personnel, as mis-targeting can lead to misclassifications. While increasing the number of electrodes may improve targeting accuracy, it also raises concerns about computational demands and real-time processing requirements. Additionally, intramuscular systems introduce substantial challenges beyond electrode placement. Their invasive nature increases the risk of infection, electrode migration, and potential tissue damage over time. Surgical implantation and the routing of leads present technical challenges, while transcutaneous connectors pose risks related to physical stability and long-term biocompatibility. It is also important to acknowledge that our study focused on well-controlled isometric contractions, where signal decomposition is highly reliable, but the system remains untested in dynamic, naturalistic conditions where movement alters action potential shapes.

In conclusion, we have proven the possibility of decoding the behaviour of the pools of motoneurons that physiologically innervate some of the muscles responsible for finger contractions. This decoding was demonstrated in two healthy individuals with three implanted electrodes each and was shown to provide information that can be used for highly intuitive and accurate control commands. The accuracy of control derived from motoneurons extracted from HD-iEMG was near-perfect for both participants and was successfully tested across a range of conditions, encompassing six to 16 distinct fine finger tasks. 

\backmatter

\section{Methods}\label{sec11}

\subsection{Data Acquisition}

\subsubsection{Participants}
Two healthy men, aged 39 and 30 years, were recruited for this study. Both participants had no history of neurological or musculoskeletal disorders that could affect motoneuron functionality. The small participant sample was due to the relatively complex implant procedure, which included three microelectrode arrays per participant, with ultrasound-guided insertion. All experimental procedures adhered to the ethical guidelines set by Imperial College London. The studies were performed according to the Declaration of Helsinki, with an informed consent form signed by all participants before each experiment (ICREC Project ID 19IC5640 for the intramuscular recordings; JRCO Project ID 18IC4685 for the MRI acquisition).

\subsubsection{Experiments} \label{methods: experiments}
HD-iEMG signals were recorded from forearm muscles using multi-channel intramuscular electrodes designed for acute recordings~\cite{muceli2022blind, muceli2015} (Fig. \ref{fig: electrode_configs}A). Additionally, three surface electrode grids (IED 4mm) with 64 recording sites each (Fig. \ref{fig: electrode_configs}C) were positioned over the targeted muscles to concurrently acquire HD-sEMG signals. This configuration enabled concurrent measurement of the electrical activity both within the muscle and at the skin surface. The EMG signals were acquired using a multi-channel amplifier (OT-Bioelettronica, Torino, Italy). All EMG data were sampled at 10,240 Hz, and digitised with 16-bit resolution. The HD-EMG signals were bandpass filtered (Butterworth, order 5) between 100-4400 Hz (HD-iEMG) and 20-500Hz (HD-sEMG). Recordings were made in single-differential configuration with a reference electrode placed on the wrist.

The participants performed force-tracking tasks following trapezoidal force trajectories at 15 \% of their MVC (Fig. \ref{fig: data}B). These tasks were executed individually for each finger (thumb, index, middle, ring, and little finger) in flexion and extension. Additionally, forces were exerted by combinations of fingers (thumb and index; index and middle; middle and ring; index with middle and ring; thumb with index, middle, ring, and little finger) in flexion and extension, following the same trapezoidal profile. As shown in Figure \ref{fig: setup}, the fingers were constrained within the flexion and extension load cell, ensuring no dynamic movement between states was involved. During all tasks, participants maintained isometric contractions, focusing on the specified contractions. Each trial consisted of a 1-second initial rest period, followed by a 3-second ramp phase (5 \%MVC/s), a 20-second constant-force hold, and a ramp-down phase (5 \%MVC/s), concluding with a 1-second rest period. A 2-minute rest was provided between contractions to minimise fatigue. Two repetitions were recorded for each task.

A total of 20 types of contractions (classes) were performed. In the following, these classes were labelled as follows: thumb flexion (T-F), index flexion (I-F), middle flexion (M-F), ring flexion (R-F), little flexion (L-F), thumb and index flexion (TI-F), index and middle flexion (IM-F), middle and ring flexion (MR-F), index and middle and ring flexion (IMR-F), thumb and index and middle and ring and little flexion (TIMRL-F) thumb extension (T-E), index extension (I-E), middle extension (M-E), ring extension (R-E), little extension (L-E), thumb and index extension (TI-E), index and middle extension (IM-E), middle and ring extension (MR-E), index and middle and ring extension (IMR-E), thumb with index, middle, ring, and little extension (TIMRL-E).

We classified the different tasks with a series of analyses. First, we attempted the classification of 16 classes that included all individual fingers in flexion and extension, and three combinations of two fingers (T-F, I-F, M-F, R-F, L-F, IM-F, MR-F, IMR-F, T-E, I-E, M-E, R-E, L-E, IM-E, MR-E, IMR-E). Then, we focused on the tasks mostly associated to the targeted muscles, which resulted in 12 classes: I-F, M-F, R-F, IM-F, MR-F, IMR-F, I-E, M-E, R-E, IM-E, MR-E, IMR-E. Finally, as a proof of concept of a specific application of the mapping system - the control of a prosthetic hand - we selected tasks that are functionally important for amputees and challenging to decode with state-of-the-art methods~\cite{kashiwakura2023task, jang2011survey}. This group of tasks included power grip (TIMRL-F, TIMRL-E), pinch (TI-F, TI-E), and individual index/thumb finger activation (T-F, T-E, I-F, I-E), for a total of eight classes. All tasks were performed isometrically, with the corresponding fingers activated simultaneously.
However, the power grip class was excluded for participant one due to the inability to correctly perform the coordinated contractions of the thumb, index, middle, ring, and little fingers. Specifically, participant one exhibited oscillations in the force applied by the little finger, with fluctuations of amplitude of around 7 \%MVC. These force variations led to inconsistent activation of the motoneurons, which were subsequently removed by the automatic cleaning process. As a result, the last analysis resulted in six classes for participant one (T-F, I-F, TI-F, T-E, I-E, TI-E) and eight for participant two (T-F, I-F, TI-F, TIMRL-F, T-E, I-E, TI-E, TIMRL-E).

\subsubsection{Instrumented platform}
A finger force measurement platform was designed to be adaptable to different hand sizes, adjustable in wrist angle, and equipped with high-resolution and cost-effective load cells to capture force signals. A 10-mm polycarbonate board was used for the platform body because of its light weight. Precision laser-cutting of load cell rails was executed using CAD software. The platform features ten 10-kg TAL220 load cells (two per finger, one for flexion, and one for extension), selected based on the maximum observed male exertion forces of 7-8 kg~\cite{carabello2022novel, cort2011maximum}. The load cells are mounted on adjustable rails with rolling pins and connected to 3D-printed mounts to accommodate varying finger thicknesses. The EMG signals and the force data were acquired simultaneously by the Quattrocento amplifier. Force signals were amplified using Forza pre-amplifiers linked to the amplifier. A 3D-printed holder housed the 10 pre-amplifiers, optimising the setup for accurate force measurement across various hand sizes.

\subsection{Classification}
We examined the accuracy in classifying intended motor tasks into predefined categories by comparing two approaches: one that used decomposed motoneuron activity (i.e., the timings of motoneuron discharge events) and another that used global features derived from the interference signal (e.g., signal amplitude).

To classify tasks using motoneuron behaviour, the HD-iEMG signals were decomposed into the contributing motoneurons. The filters associated with each source (motoneuron) were aggregated into a comprehensive dictionary of filters for each task. Figure \ref{fig: classification_pipeline} panel A provides a schematic representation of this filter bank. In the diagram, each colour corresponds to a different class label, indicating the specific task being represented. The width of each bar reflects the number of filters associated with that task, illustrating how many motoneuron filters contribute to the classification of each specific task. The zoomed-in section within the magenta box offers a closer look at a representative filter from the pooled filter bank. This filter visualises the amplitude of the motoneuron across different channels, providing insights into how the neural activity is distributed spatially across the electrode array. Conversely, for global EMG classification, the root mean square of the EMG signals served as the feature for a C-Support Vector Machine classifier (Fig. \ref{fig: classification_pipeline}C). To mimic a real-life mapping system, the calibration phase was conducted using the first recorded repetition, and the testing phase was performed using a repetition recorded later in the day.

\subsubsection{Finger task classification based on motoneuron behaviour}
Motor unit spike trains were extracted through blind source separation of HD-iEMG recordings obtained during isometric finger tasks at 15 \%MVC (see Appendix \ref{secA0} for more details on EMG decomposition). The tasks included individual finger flexion and extension as well as multi-finger combinations performed isometrically. The decomposition process produced task-specific sets of motoneuron filters, which were pooled to create a comprehensive bank of motoneuron action potential filters $\pmb{\mathbb{F}}_i \in \mathbb{R}^{N \times M}$, where $N$ represents the total number of extracted motoneuron filters and $M$ the number of EMG channels. Each motoneuron filter $\pmb{\mathbb{F}}_i$ in this set was labelled according to the specific finger movement task it was derived from (e.g., thumb flexion corresponding to class 0).

These reference motoneuron filters from the 15 \%MVC contraction were then applied to the multi-channel iEMG signals $\mathbf{E}j \in \mathbb{R}^{M \times D_R}$, where $D_R$ is the duration of the recording, recorded during a separate 15 \% MVC trial involving all fingers performing both flexion and extension contractions. This produced the corresponding discharge timings, represented as IPTs $\mathbf{I}{ij} \in \mathbb{R}^{N \times D_R}$ for each motoneuron $i$ and task $j$. Each IPT $\mathbf{I}_{ij}$ was clustered to separate the peaks (i.e., the source) from the noise components. Only motoneurons with robust clustering quality (silhouette value \textgreater 0.9), physiological discharge periodicity (coefficient of variation \textless 35\%), and firing rates within the physiological range for isometric contractions (mean discharge rate \textless 30 Hz) were retained for further processing. The discharge timings of these units were temporally segmented into 100-ms time bins with 50\% overlap. This window was selected to be long enough for reliable classification while remaining short enough to minimize delay in real-time control applications. The windowing resulted in a binary matrix $\mathbf{B} \in \mathbb{R}^{N\times K}$, where $K$ represents the number of bins. For each time bin $k$, the corresponding class label was assigned according to the filter $\pmb{\mathbb{F}}_i$ that produced the source, creating a final label vector $\mathbf{L} \in \mathbb{R}^K$. If no predictions were made for a given bin (i.e., no motoneurons were active), it was labeled as N/A. Finally, for each time bin, the class with the majority of votes was assigned to that bin. If there were no predictions overall for a bin, it was labeled as N/A and counted as a misclassification in the final accuracy calculation. The method, therefore, relies on unique motoneurons being consistently active throughout the entire task. Figure \ref{fig: classification_pipeline}A,B shows a schematic of the method.

\subsubsection{Finger classification based on global EMG}
For myoelectric decoding of dexterous hand tasks, a linear SVM classifier with an L2 regularisation penalty factor of 0.1 was employed (Fig. \ref{fig: classification_pipeline}C). The SVM classifier is a widely used method in EMG pattern recognition~\cite{oskoei2008support, purushothaman2018identification, dhindsa2019performance}.
The HD-EMG signals were bandpass filtered (Butterworth, order 5) between 100-4400 Hz (HD-iEMG) and 20-500 Hz (HD-sEMG) to remove noise. Global normalisation factors based on the maximum absolute value across the pooled training set were derived and applied to the test set for consistent scaling. Outlier channels with baseline noise exceeding 3 standard deviations from the training mean were removed from all training and testing data.
The pre-processed EMG was segmented into 100-ms intervals, with 50\% overlap using a sliding window, preparing the bins similarly to how they were organised for the classification based on motoneurons. The RMS values of the EMG were extracted as the input features. This was done in the same way for both surface and intramuscular high-density recordings.
Moreover, to simulate a lower-density iEMG electrode configuration, we selected the first and last channel for each electrode, resulting in an IED of 2 cm (Fig. \ref{fig: electrode_configs}B). The difference between the two electrodes was computed for each array, resulting in 3 feature channels. This configuration was chosen to emulate structures like the implantable myoelectric sensors (IMES) electrodes, currently being tested to control multifunction hand prostheses~\cite{weir2008implantable, merrill2011development, salminger2019long}. For the surface electromyography (sEMG), we selected eight channels in the central column of the electrode grid, mimicking the configurations found in commercial low-density systems such as the MYOband (Fig. \ref{fig: electrode_configs}D).

\section{Acknowledgments}
We thank the participants who generously contributed their time and effort to this study. We also thank Fabio Bolognesi for his help with the synchronization of the force and EMG recordings.

A.G. was supported by UK Research and Innovation [UKRI Centre
for Doctoral Training in AI for Healthcare grant number EP/S023283/1] and by Huawei Technologies Research \& Development (UK) Limited. J.I.P. was supported by project ECHOES (ERC Starting Grant 101077693) and by a Consolidación Investigadora grant (CNS2022-135366) funded by MCIN/AEI/10.13039/
501100011033 and UE’s NextGenerationEU/PRTR funds. S.M. was supported by Chalmers Life Science Engineering Area of Advance. A.K. was supported by NISNEM (EPSRC EP/T020970/1). D.F. was supported by NaturalBionicS (ERC Synergy 810346). 

\section{Conflict of interest}
N/A

\section{Ethics approval and consent to participate}
All experimental procedures adhered to the ethical guidelines set by Imperial College London. The studies were performed according to the Declaration of Helsinki, with an informed consent form signed by all participants before each experiment (ICREC Project ID 19IC5640 for the intramuscular recordings; JRCO Project ID 18IC4685 for the MRI acquisition).

\section{Data availability}
The data that support the findings of this study are available from the corresponding author (D.F.) upon reasonable request.

\section{Code availability}
The code developed to analyse the data is available from the corresponding author (D.F.) on request.

\section{Author contribution}
A.G., J.I.P., S.M., D.F. conceptualized the study. A.G., J.I.P., A.K., S.M., D.F. performed the experimental measures. A.G. performed the data analysis. F.B., E.D., G.I. provided guidance to the project. A.G., D.F. prepared the first draft of the manuscript. All authors edited the manuscript for important scientific content and all approved the final version.

\section{Consent for publication}
N/A

\bibliography{references}

\clearpage
\begin{appendices}

\section{Decomposition} \label{secA0}
The EMG signals were decoded into the discharges of the corresponding motoneurons via a blind source separation algorithm~\cite{grison2023particle}.

Multi-channel EMG signals represent combined mixtures of motoneuron spike trains. Each recorded signal, i.e., each EMG channel, is the summation of the convolution of unknown finite impulse response filters, corresponding to motor unit action potentials, with their respective sources, i.e., the sequences of motoneuron discharge timings. To make the deconvolution (i.e., \emph{EMG decomposition}) possible, this combined mixture can be converted into a linear instantaneous mixture by extending the sources to include the original \(N\) sources and their delayed versions, with delays ranging from 1 to the filter length \(L\). Similarly, the \(M\) original observations are extended by a factor \(R\) to ensure there are more observations than sources~\cite{negro2016multi}.

For each time sample $(k)$, indicating the sources, observations and noise, respectively, with $\underset{=}{s}(k)=\left[s_1(k), s_2(k), \ldots, s_N(k)\right]^T, \underset{=}{x}(k)=\left[x_1(k), x_2(k), \ldots, x_M(k)\right]^T$ and $\underset{=}{n}$ $(k)=\left(n_1(k), n_2(k), \ldots, n_N(k)\right)^T$, where $T$ is the transpose operator, the extended model is as follows:
\begin{equation}\label{eq1}
\underset{=}{\widetilde{x}}(k) = [\underset{=}{\widetilde{H}} \widetilde{s}(k) + \underset{=}{\widetilde{n}}(k)] \quad \text{for} \quad k = 1, \ldots, D_R
\end{equation}
where $D_R$ is the duration of the recording in samples, and with the extended sources, observations and noise as $\widetilde{s}(k)=\left[\widetilde{s_1}(k), \widetilde{s_2}(k), \ldots, \widetilde{s_n}(k)\right]^T$, $\widetilde{x}(k)=\left[\widetilde{x_1}(k), \widetilde{x_2}(k), \ldots, \widetilde{x_M}(k)\right]^T$ and $\widetilde{n}(k)=\left[\widetilde{n_1}(k), \widetilde{n_2}(k), \ldots, \widetilde{n_N}(k)\right]^T$, where:

\begin{align*}
& \widetilde{s_j}(k) = \left[s_j(k), s_j(k-1), \ldots, s_j(k-L-R)\right] \quad & j = 1, \ldots, N \\
& \widetilde{x_i}(k) = \left[x_i(k), x_i(k-1), \ldots, x_i(k-R)\right] \quad & i = 1, \ldots, M \\
& \widetilde{n_i}(k) = \left[n_i(k), n_i(k-1), \ldots, n_i(k-R)\right] \quad & i = 1, \ldots, M
\end{align*}

The mixing matrix representing the filters is:
$$
\underset{=}{\widetilde{H}}=\left[\begin{array}{ccc}
\widetilde{h_{11}} & \cdots & \widetilde{h_{1 N}} \\
\vdots & \ddots & \vdots \\
\widetilde{h_{M 1}} & \cdots & \widetilde{h_{M N}}
\end{array}\right]
$$
with:
$$
h_{i j}=\left[\begin{array}{cccccc}
h_{i j}[0] & \cdots & h_{i j}[L-1] & 0 & \cdots & 0 \\
0 & \ddots & \ddots & \ddots & \ddots & \vdots \\
\vdots & \ddots & \ddots & \ddots & \ddots & 0 \\
0 & \cdots & 0 & h_{i j}[0] & \cdots & h_{i j}[L-1]
\end{array}\right]
$$
where $h_{i j}$ the action potential of the $j^{th}$ motoneuron recorded at the channel $i$. The noise in model \ref{eq1} includes both electronic interference and the activity of motoneurons at the skin surface, depicted by low-energy action potentials that cannot be separated.

The linear instantaneous model in Equation \ref{eq1} is inverted to retrieve the matrix of extended and spatially whitened sources. The inversion of the matrix is performed with an iterative gradient descent optimisation procedure with a contrast function that maximises the independence, or sparseness~\cite{negro2016multi}, of the estimated sources. This method extracts the motoneuron discharge timings associated with individual motoneurons and the unique representation of the associated multi-channel action potentials. The estimated sources are trains of delta functions centered at motoneuron activation instants, with varying amplitudes due to the estimation process. To separate the discharge timings from the baseline noise, K-Means clustering is applied to the estimated sources, comparing each candidate peak with the surrounding peaks.

Several blind source separation approaches have been proposed in the literature~\cite{muceli2022blind, jackel2012applicability, leibig2016unsupervised, buccino2018independent, chen20242cfastica, chen2018automatic}. However, these methods often struggle to separate sources when their waveforms are similar, resulting in overlapping signals. Current contrast functions and non-linear metrics are not optimised to resolve these overlaps. We thus utilised a novel algorithm, Swarm-Contrastive Decomposition ~\cite{grison2023particle, grison2025unlocking}, that addresses these limitations by using an adaptive contrast function that enhances spike sorting accuracy by adapting to the unique characteristics of spiking distributions.

\clearpage
\section{Lower-density configurations}\label{secA1}

\begin{figure}[!htbp]
    \includegraphics[width=\columnwidth]{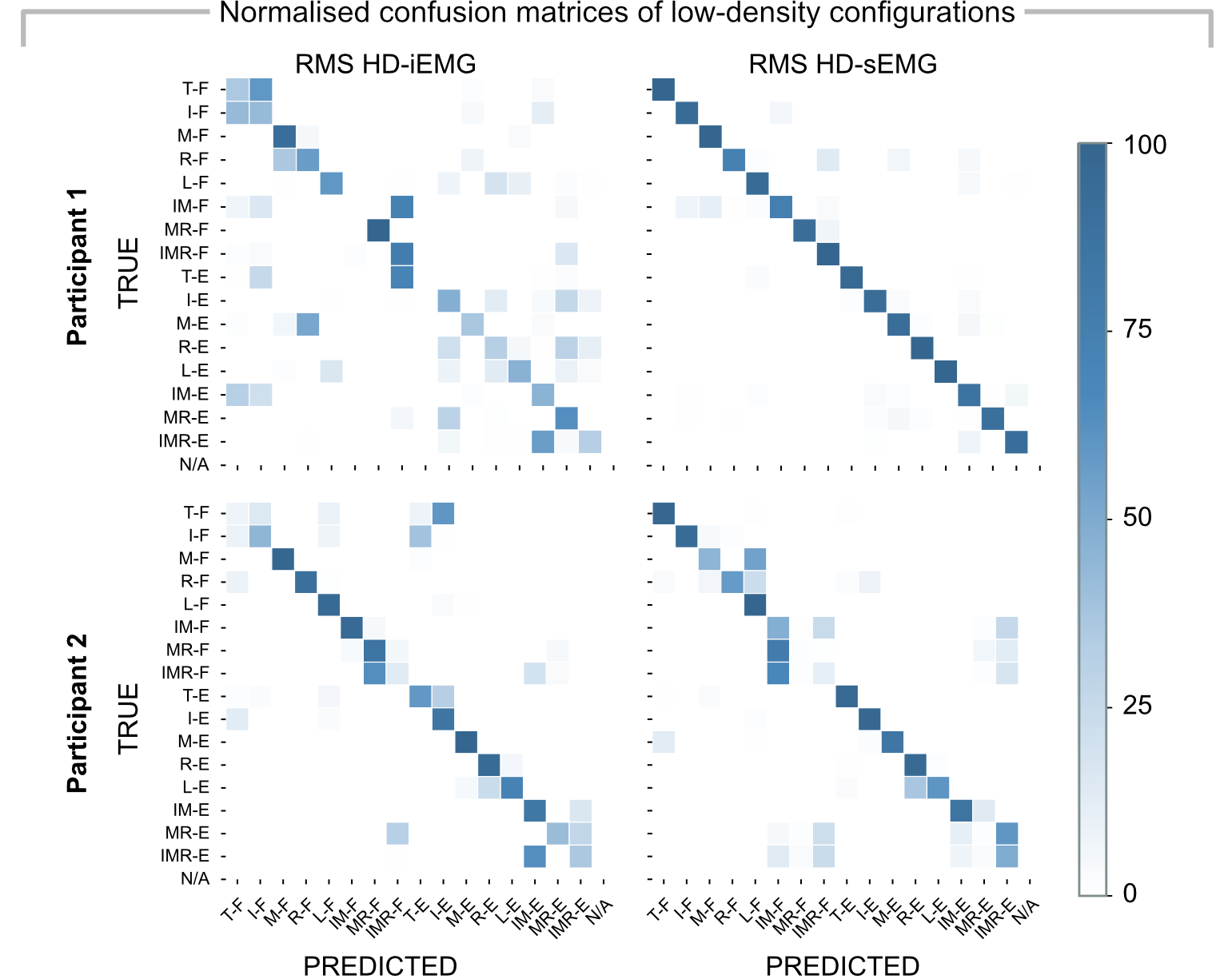}
        \caption{Normalised confusion matrices illustrating the percentage of predictions for each of the 12 flexion and extension tasks, comparing true labels (y-axis) with predicted labels (x-axis). The matrices show predictions for participant one on the first row and participant two on the second row, across different data types: RMS values of simulated bipolar (first column), RMS values of 8-channel sEMG (second column). }
        \label{fig: bipolar_surf_16}
\end{figure}

\begin{figure}[!htbp]
    \includegraphics[width=\textwidth]{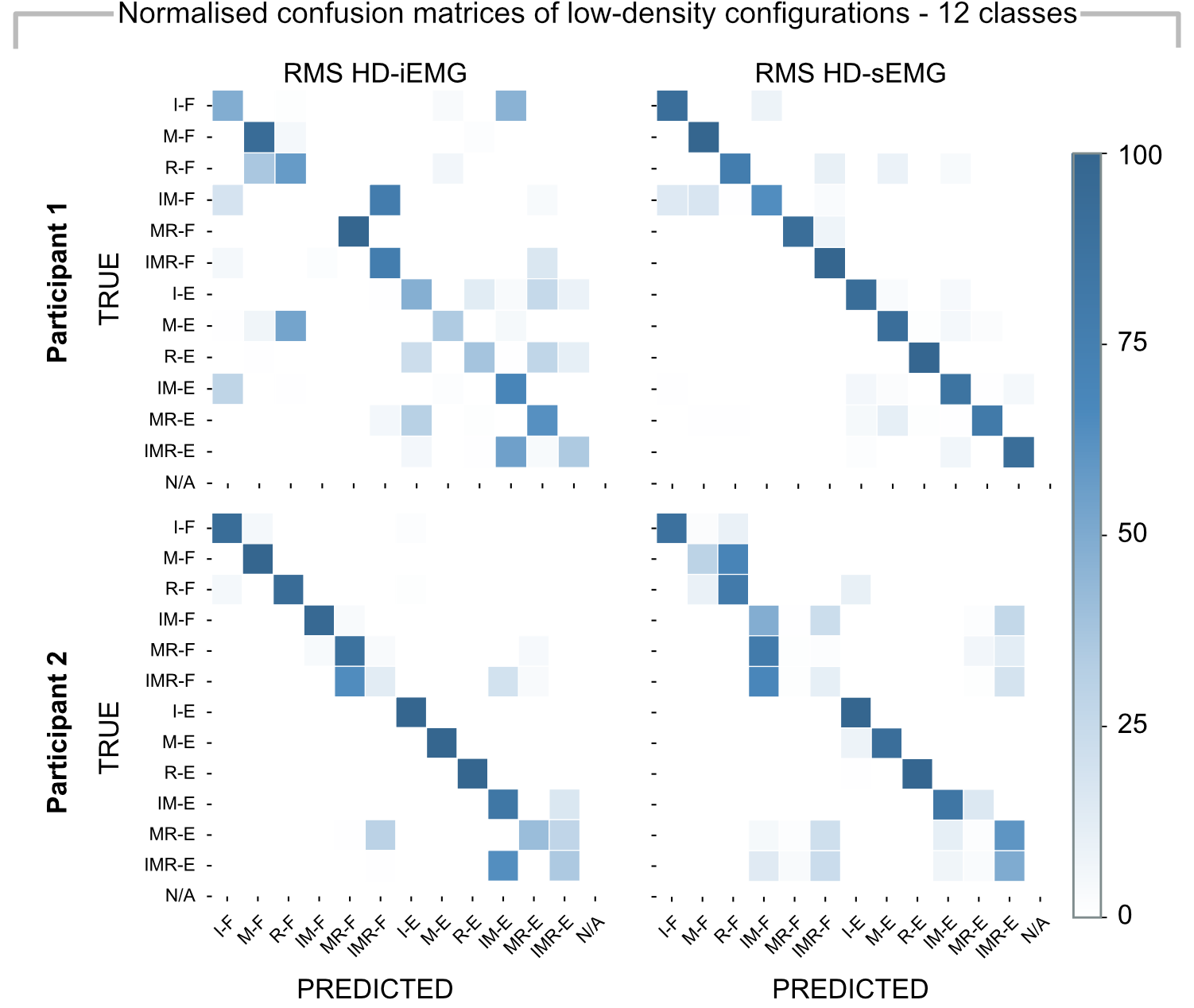}
        \caption{Normalised confusion matrices illustrating the percentage of predictions for each of the 12 flexion and extension tasks, comparing true labels (y-axis) with predicted labels (x-axis). The matrices show predictions for participant one on the first row and participant two on the second row, across different data types: time domain and RMS values of simulated bipolar (first column), time domain and RMS values of 8-channel sEMG (second column). }
        \label{fig: bipolar_surf_12}
\end{figure}

\clearpage

\end{appendices}

\end{document}